\documentclass[slac_one]{revtex4}
\usepackage{graphicx}
\usepackage{fancyhdr}
\newcommand{\be}{\begin{equation}}
\newcommand{\ee}{\end{equation}}
\newcommand{\bea}{\begin{eqnarray}}
\newcommand{\eea}{\end{eqnarray}}

\begin{document}

\title{Aspects of Unparticle Physics} 

%

\author{Arvind Rajaraman}
\affiliation{Department of Physics and Astronomy, University of California,
Irvine, CA 92697, USA}

\begin{abstract}
We review some theoretical and experimental issues in unparticle physics, focusing mainly on
collider signatures.
\end{abstract}

\maketitle

\thispagestyle{fancy}


\section{INTRODUCTION AND MOTIVATION} 
It is widely believed that the Standard Model is not a complete theory
of particle physics, and that there is a new physics sector coupled to
the Standard Model, which addresses various problems of the Standard Model such as the hierarchy problem
and the identity of dark matter.
A priori, this new physics can be of several basic forms. It can be
weakly coupled, like  supersymmetric theories. Another class of theories is technicolor,
where the new physics is
strongly coupled.

Logically, there is a third option; the new physics could be conformal (or scale-invariant).
This is the idea of unparticles. The idea (and the name) was introduced by  Georgi in
~\cite{Georgi:2007ek} (for a recent review of unparticles, and further
references to the literature, see \cite{review}).
As we shall discuss here,
unparticles  may provide many interesting and unexpected signals at the upcoming Large Hadron Collider (LHC).

\section{THE UNPARTICLE THEORY}

To discuss the signals of unparticles, we need
\begin{itemize}
\item A description of the hidden sector (the CFT)

\item A form for the interactions between the CFT and
     the SM.

     \end{itemize}

\subsection{The hidden sector}
There are many possible examples of conformal theories, e.g. Banks-Zaks theories
\cite{Banks:1981nn}
and supersymmetric QCD in the conformal window\cite{Intriligator:1995au}.
We will take a very  simple model for the conformal sector by focusing on a few
operators of the theory rather than the full complexity. In fact,
we will assume that the only coupling to the
conformal sector is through a single operator $O_{ U}$ (
this operator $O_{U}$ can be a scalar, vector,
fermion or a higher spin object.)

Now the propagator for this operator is fixed by conformal invariance.
For example, the propagator for scalar unparticles in an exactly conformal theory
is fixed to be
~\cite{Georgi:2007si,Cheung:2007zza}
\begin{equation}
   T\langle 0 |O_{ U} (x) O_{ U}^\dagger (0)|0 \rangle =   i B_d (p^2)^{d-2}
\end{equation}
where
$B_d$ is a constant,
and $ d$ is the dimension of the operator $O_{U}$.
There are similar expressions for vector and fermion operators.

The momentum space  propagator can also be written as a
dispersion integral~\cite{Georgi:2007si,{Stephanov:2007ry}}
\bea
   T\langle 0 |O_{ U} (x) O_{ U}^\dagger (0)|0 \rangle =
i {A_d\over 2\pi}\int_0^\infty dM^2 {\rho(M^2)\over p^2-\mu^2-M^2+i\epsilon}
\eea
with $\rho(M^2)=(M^2)^{d-2}$.
This representation shows that the propagator can be understood
as a sum over resonances where the resonance
masses are  continuously distributed.  In particular, there is no mass gap.

\subsection{Constraints on the dimension}
The dimension is constrained by the unitarity of
the conformal algebra\footnote{These limits can be avoided if the theory is
scale-invariant, but not conformal.}. The unitarity
imposes lower bounds on the dimension \cite{Mack:1975je}
\begin{itemize}
\item Scalar operators:       $d\ge 1$
\item Fermion operators:   $d\ge 3/2 $
\item Gauge invariant vector operators:       $d\ge 3$
     \end{itemize}

On the other hand we can also find {\it upper} bounds on the dimension.
Since $\rho(M^2)=(M^2)^{d-2}$. we see that for $d > 2$, the
theory is ultraviolet sensitive; small changes at high energies
radically alter the propagator. This has several consequences.
In particular,
we find singular behavior in many situations with $d > 2$ e.g. the energy density at finite
temperature and some
cross-sections are proportional to (2-d)~\cite{Cacciapaglia:2007jq}. These imply that
we should restrict ourselves to $d < 2$ ($d < 5/2$ for fermions).

Combining this with the lower bounds above, we find that vector and
higher spin operators are problematic; accordingly we will focus on
scalar operators.

\subsection{Standard Model Interactions}

The unparticle operator  is coupled to the Standard Model by
terms of the form $O_{ U} O_{SM}$.
The only constraint on these couplings is dimensional analysis. There are
therefore a
huge number of possible couplings.
Schematically, we can have a Lagrangian
\bea
L_{int} =       \Lambda^{2-d} O_{ U} H^2
+c_{\psi}\Lambda^{1-d} O_{ U} \bar{\psi}\psi
+c_F\Lambda^{-d} O_{ U} F^2
\eea

For every quark pair, lepton pair and gauge field,
we have an independent coupling. These couplings are unconstrained by theory.

\section{The Unparticle Mass Gap}
       We now begin the study of the experimental signals of
unparticles.


As we have mentioned above, exact scale invariance precludes a mass gap for unparticles.
In an exactly scale-invariant theory,
unparticles would mediate long range forces.
They would also contribute to precision experiments
like $(g-2)_\mu$  through loop effects. Since they are
effectively massless, this can be a big effect.
Indeed, as shown in
several papers (see e.g. \cite{Cheung:2007zza,Luo:2007bq,Chen:2007vv,Ding:2007bm,Liao:2007bx}),
the  constraints are strong enough
to rule out any possibility of seeing unparticles at
the LHC.

We now argue that in fact interactions induce a mass
gap for unparticles. This removes all these low energy constraints.

This happens due to the Higgs couplings $\Lambda^{2-d} O H^2$.
This coupling is a relevant operator; it becomes important at low energies.
In particular it breaks the conformal invariance of the low-energy theory
\cite{Fox:2007sy}. This is clear from the fact that once the Higgs field gets a vev $H= v+h$, we get
linear terms as well as Higgs-unparticle mixing.
This then introduces a scale  $\mu$ with $\mu^{4-d}  =   \Lambda^{2-d}  v^2$   and  breaks scale
invariance in the hidden sector.

The introduction of the scale modifies the density of
resonances  $\rho(M^2)$. The modification is model dependent \cite{Delgado:2007dx}, but
generically, we find that a mass gap is introduced i. e.
              $\rho(M^2)=0$      for    $M^2 \ll m^2$.
Also, for high energies, the density is unchanged:
                   $\rho(M^2)= (M^2)^{d-2}$ for $M^2 \gg m^2$.

Since  $\Lambda \sim v \sim 100$ GeV,
we find that in the absence of fine tuning, the mass gap is at least
a few GeV. This immediately implies that there are no long range forces from unparticle exchange.
 Precision constraints (say from $(g-2)_\mu$ ) essentially
disappear.

 This further implies that low energy experiments
 are not sensitive to unparticles: unparticles are best probed at colliders.
 We therefore turn to collider signals of unparticles.

\section{COLLIDER SIGNATURES: UNPARTICLE DECAYS}

\begin{figure*}[t]
\centering
  \includegraphics[width=135mm]{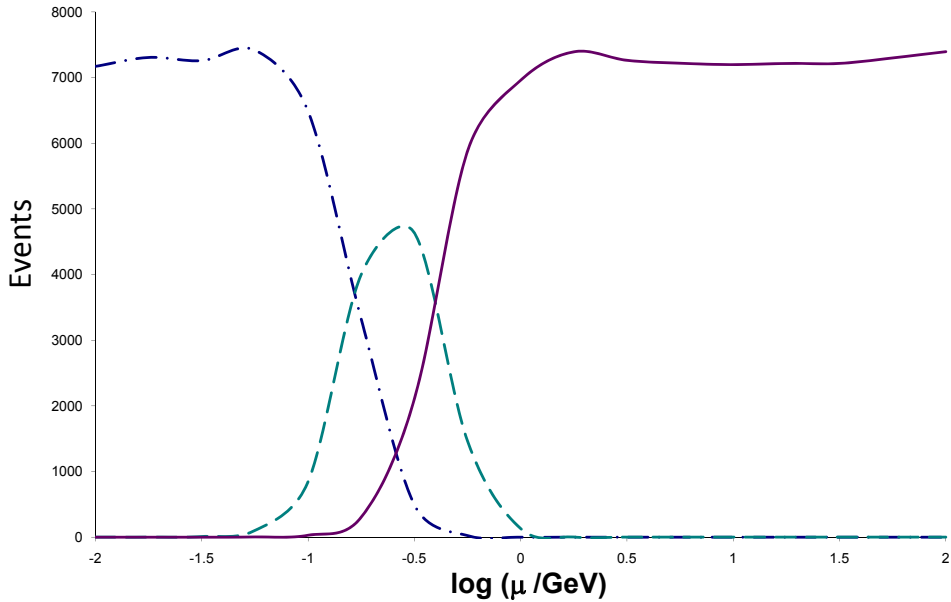}
  \caption{Number of events with $10$ fb$^{-1}$ of LHC data as a function of $\mu$.
The solid (red) line corresponds to the number of prompt events,
dot-dashed (blue) corresponds to the number of monojet events, and
dashed (green) is the number of delayed events. We have taken $d=1.1$ and
$\Lambda=10000$ GeV.}
\label{events}
\end{figure*}

For definiteness, we will henceforth consider a situation where the unparticle
couples primarily to gauge bosons
\bea
L_{int}={O_{ U}F_{\mu\nu}F^{\mu\nu}\over \Lambda_F^d}+{O_{ U}
G_{\mu\nu}G^{\mu\nu}\over \Lambda_G^d}
\label{ints}
\eea
where $F_{\mu\nu}, G_{\mu\nu}$ are the electromagnetic and color field
strengths respectively, and
$\Lambda_F, \Lambda_G$ are scales parametrizing the
couplings. We will take for simplicity
 $\Lambda_F\sim \Lambda_G=\Lambda$.

Unparticles can now be produced at hadron colliders
through processes like  $gg  \rightarrow  gO_U$.
If the unparticle does not decay we get monojets
(and more generally,  missing energy signals.)
In fact, as we now show,  unparticles can decay to SM particles.
This will modify this set of signals.

The basic point is that loop corrections can modify the unparticle
propagator.
Resumming the contributions,
we obtain the full propagator
\bea
  T\langle 0 |O_{ U} (x) O_{ U}^\dagger (0)|0 \rangle = {iB_d\over
(p^2-\mu^2)^{2-d}-B_d\Sigma(p^2)}\nonumber
\eea
where the loop diagram is $-i\Sigma(p^2)$. The
propagator can develop an imaginary piece and this leads to unparticle decay.
 The width  is proportional to the imaginary
part of $B_d\Sigma(p^2)$ .

If unparticles are produced at colliders, they can
themselves decay back to Standard Model particles.
Depending on the lifetime, we have different signals. If the lifetime
is   short, we have     prompt decays. If the    lifetime is      very long,
we get      monojets and missing energy signals.
Most interesting is the  intermediate situation, where we can get
       delayed events and displaced                  vertices.

We calculate the number of each type of event,
with 10 fb of LHC data. We take d=1.1,
 $\Lambda$ = 10 TeV. We require the gluon jet to have energy $> 100$ GeV.
We shall further assume that
the detector is ~ 1m in size, and that  delays of 100ps can be measured \cite{Vigano:2006zz}.

    Our results are shown in Fig \ref{events}. For $\mu > 10$ GeV, we  only
    have  prompt events.
There are a significant number of monojets only if  $\mu < 100$ MeV, which requires some
fine tuning as discussed earlier.
Interestingly, in an
intermediate range  ($\mu \sim $ 1 GeV), there are a large number of
delayed events. These could have a striking signal at the LHC.


\begin{figure*}[t]
  \includegraphics[width=135mm]{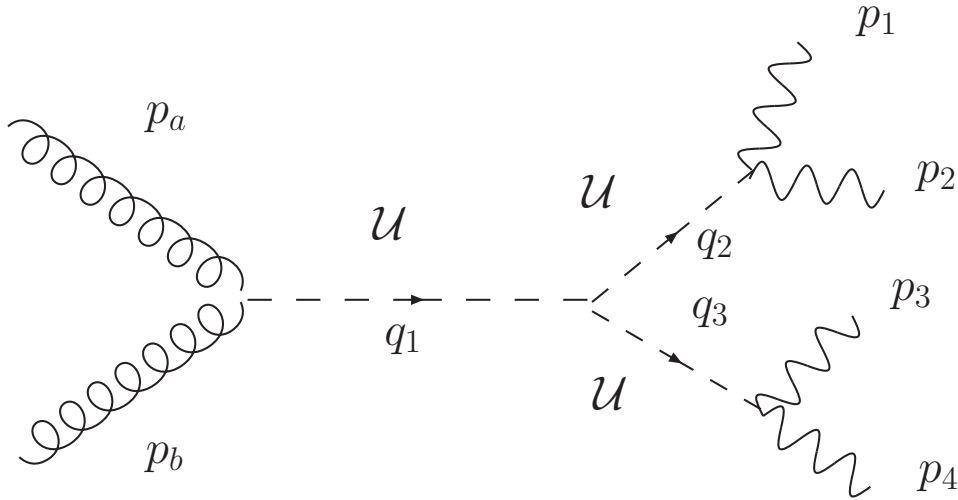}
  \caption{The process $g g \to \gamma \gamma \gamma \gamma$ mediated by
  unparticle self-interactions.}
  \label{feynmanfig}
\end{figure*}


\section{Unparticle Self-Interactions}

We now look at signals from unparticle self-interactions.

The three-point interactions of operators are  fixed by
conformal invariance up to an overall constant
\begin{equation}
   \langle 0 |O (x)\, O (y)\, O^\dagger (0)|0 \rangle =
 \frac{C}{|x - y|^d \, |x|^d \, |y|^d}\, ,
\end{equation}

Assume again that the unparticle couples to gluons
and photons
as in eq. (\ref{ints}). At the Tevatron and LHC
we can now have four-photon processes as shown in Fig \ref{feynmanfig}.
\begin{figure}[t]
  \includegraphics[width=135mm]{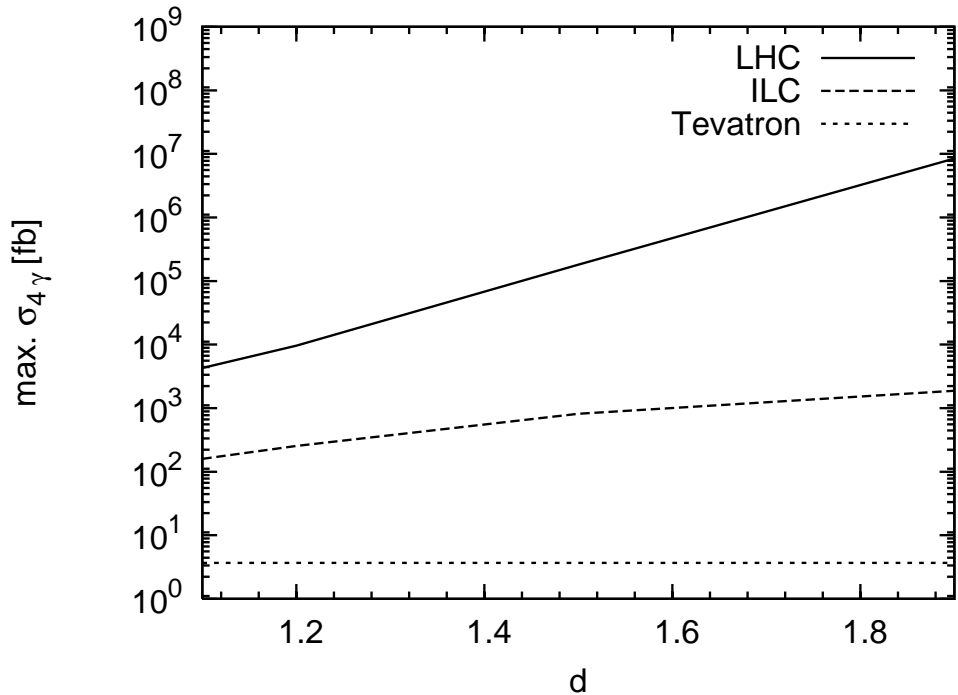}
  \caption{Cross-sections at the LHC and ILC for the process $g g \to \gamma \gamma \gamma \gamma$.}
  \label{rate}
\end{figure}

The rate of these processes is controlled by the constant $C$ in the three point-function, which
 is not constrained by theory. Now we can use experimental input to constrain the
 allowed value of $C$. This analysis was performed in \cite{Feng:2008ae}, where it
 was found that the primary bound comes from
the nonobservation of excess four-photon events at the Tevatron.
Assuming that $C$ is as large as the experimental limit, we can predict the rate of these processes at
the LHC and ILC. These rates
are shown in Fig \ref{rate}. We find that there are huge rates for these processes,
which could again provide striking signals
of unparticles.

To conclude,
we have shown that
unparticles are an interesting alternative model for
the hidden sector. In these models, we
can obtain many unique and striking signals at colliders such as
delayed events, displaced vertices and multiphoton processes.
Further work on unparticles may well lead to other new and interesting signals
that would otherwise have been missed.

\begin{acknowledgments}
This work is supported in part by
NSF Grants No.~PHY--0354993 and PHY--0653656.
\end{acknowledgments}

\end{document}